# Stability of graphene hyperbolic pseudospheres under harsh conditions

T. P. C. Klaver[1*], R. Gabbrielli[2], V. Tynianska[3], A. Iorio[3,4,5], D. Legut[1,6]

[1] *IT4Innovations, VSB-Technical University of Ostrava, 17. listopadu 2172/15, 708 00 Ostrava, Czech Republic*

[2] *independent researcher, Italy*

[3] *Institute of Particle and Nuclear Physics, Faculty of Mathematics and Physics, Charles University, V Holešovičkách 2, 18000 Prague 8, Czech Republic*

[4] *Department of Physics, University of Calabria, 87036 Rende (CS), Italy*

[5] *Democritos Technology, Rybna, 110 00 Prague 1, Czech Republic*

[6] *Department of Condensed Matter Physics, Faculty of Mathematics and Physics, Charles University, Ke Karlovu 3, 121 16 Prague 2, Czech Republic*

We demonstrate the high stability of simulated graphene hyperbolic pseudospheres under large externally imposed deformations and high temperature annealing. Hyperbolic pseudospheres are produced in a two-step Molecular Dynamics simulation process. First, carbon atoms are forced down a thin three-dimensional volume of a chosen shape. During this extrusion process the carbon atoms form a precursor to graphene that is unrealistically less stable than graphite or diamond. Then the unstable carbon structure is annealed inside the thin volume at high temperature, turning the carbon into realistic polycrystalline, curved graphene. Point defects naturally appear in numbers and places that stabilize the graphene in the desired shape, without high residual stresses. We applied this new methodology to the creation of graphene hyperbolic pseudosphere surfaces, which reproduce analogs to some aspects of classical or quantum gravity. The free edges of the pseudosphere cause bending of the graphene. When these free edges are removed from the simulations by attaching periodic flat graphene sheets to the pseudosphere edges, the carbon atoms assume positions just some tenths of Å from the mathematical hyperbolic pseudosphere surface. In demanding tests of their stability, the hyperbolic pseudospheres proved stable against 20° shearing or 20% elongation and then being released, which eventually raised their temperatures by ~300 K. Our methodology is relatively easy to use and offers a practical way to create simulated curved graphene surfaces of almost any shape. It allows for thorough testing in advance of the stability of graphene shapes that are to be produced experimentally.

---

[*] contacting author, klaver2@gmail.com

1 Introduction

Motivated by the possibility to reproduce analog-gravitational phenomena, such as the Hawking Effect on graphene [1-5], some of us previously reported the *in silico* construction of Beltrami pseudospheres (BELs) out of carbon [6].
The theoretical framework of [6] is based on the massless Dirac field behavior of the low-energy/long-wavelength conductivity electrons of graphene [7]. As such the carbon system enjoys a generalized scaling symmetry, Weyl symmetry [1], through which the boundary of a BEL can be mapped to a black hole (BH) horizon, in a limit [4]. This result ignited further research, such as the exploitation of graphene as an analog system able to reproduce various aspects of classical and quantum gravity [8]. However, phenomena that require an event horizon are only within reach in a limit, namely for very large carbon BELs [4]. Hence a great number of atoms in the lattice and difficult to handle length-to-width ratio are necessary. In fact, such structures have not yet been realized in laboratories, despite the efforts with simulations [6] and preliminary experimental attempts. Therefore, it is of great importance to search for easier-to-build carbon structures on which analog gravitational phenomena that require an event horizon may still take place.
In this paper we take the first steps in that direction, by moving from the Weyl-symmetry approach, that necessarily requires the above mentioned limiting process, to a novel time-slice approach, that, when the pseudosphere is a hyperbolic pseudosphere (HYP), no longer requires the limit. Notice the unique role played by HYP, among the three pseudospheres of classic differential geometry, the elliptic (ELL), the BEL and the HYP, see, e.g., [9, 10].

The general theoretical strategy of the time-slice approach will be discussed in detail in forthcoming papers [11, 12]. Here the focus is on the experimental feasibility of HYP, its structural properties and stability, and resilience under experimental conditions.

About the theoretical set-up, let us just mention here, though, that first of all we are in a (2+1)-dimensional system, with a 2-dimensional membrane (graphene) and time. In this context the BH to consider is the famous Bañados-Teitelboim-Zanelli (BTZ) BH [13]. The time-slice approach involves the freezing of the BH time, so that we only retain the spatial part of the metric. When one does that, each one of the three pseudospheres, ELL, BEL and HYP, corresponds to specific different phase of the evolution, with HYP being the only one corresponding to a proper BH with nonzero mass, see [11] and later in this paper.

This identifies the finite size carbon HYP as the carbon structure that may reproduce important analog gravity. Indeed, the location of the general relativistic horizon is not reached in a limit, as it happens for BEL in the Weyl approach, but instead it is right there, in the middle of the finite HYP structure. No limit to an infinitely large structure is required. For this fact to play the desired general relativistic role, the massless nature of the (pseudo-)relativistic Dirac quasiparticles is crucial (as it is for scale/Weyl symmetry [1]). With these motivations we produce here a carbon HYP in a simulation laboratory, to test in advance if its mechanical stability is sufficient to conduct experiments on it.

The pseudosphere construction adopted in [6] was based on surface-constrained energy minimization of a system of points interacting via a pair-wise Lennard-Jones potential where points where initially randomly scattered on the BEL surface, subsequently relaxed and finally translated into a dual configuration of atoms through a Voronoi construction. Here we take a very different approach, using computational materials science, and specifically atomistic simulation, to create a carbon HYP. We use production methods and physical processes such as extrusion, annealing, equilibration through diffusion and elimination of point defects, and stress relief, as well as some artificial, unphysical tricks that atomistic simulators conveniently have at their disposal, to produce a carbon HYP. The rather simple (and arguably crude) nature of our method offers the advantage of being easily and widely applicable. For example, apart from producing a HYP as reported in this work, our method was also used to easily produce a BEL pseudosphere. It could also be used to produce graphene bodies with a wide variety of other shapes, including ones for which no analytical description is available, for which the number and positions of required defects cannot be theoretically determined, and which might include more complex defects than just (combinations of) pentagons, heptagons and octagons. Another advantage of our method is that recrystallisation and rearrangement of defects during annealing relieves most stress from the formed graphene, giving the surfaces high mechanical stability. This allows testing of graphene surfaces for real world practical applications, such as in desalination filters [14]. The buildup of this paper is as follows. In section 2 we first briefly present some of the relevant mathematics of HYPs. In section 3 we present computational details of how our method works and how it can be applied to easily produce a wide range of graphene surfaces of almost any desired shape. In section 4 we present results of how simulated nano-scale extrusion can produce a highly unstable precursor to a carbon HYP, how relaxation and annealing then turn the unstable precursor into realistic, mechanically stable curved graphene, and how attaching flat graphene sheets to the HYP can help it retain its shape. Finally in section 4, we show how the combined system of HYP plus periodic graphene sheets is stable even under very large imposed mechanical deformation (20° shearing or 20% elongation) followed by release, as well as during high temperature annealing, and thus is feasible for experimental preparation and testing. Section 5 gives a summary of the results obtained and offers some future prospects.

2 Hyperbolic pseudospheres

The HYP is one member of the family of three pseudospheres, along with BEL and ELL, that are surfaces of revolution with constant, negative Gaussian curvature, $K = -1/r^2$. Its Cartesian coordinates are expressed in terms of the meridian coordinate, u, and of the angle, v, of the surface, as any other surface of revolution, i.e.

$$x(u,v) = R(u)\cos v \tag{1}$$

$$y(u,v) = R(u)\sin v \tag{2}$$

$$z(u) = \int^u \sqrt{(1 - (R(w)')^2)}\, dw \tag{3}$$

where $R(u)$, with its range and domain, fully identifies the surface. In our case

$$R(u) = c \cosh\left(\frac{u}{r}\right) \tag{4}$$

thus

$$z(u) = \int^u \sqrt{1 - \left(\frac{c}{r} \sinh \frac{w}{r}\right)^2}\, dw \tag{5}$$

and

$$u \in \left(-r \operatorname{arcsinh}\left(\frac{r}{c}\right),\, +r \operatorname{arcsinh}\left(\frac{r}{c}\right)\right)$$

where r is the square root of $-K$, $c = R_{min}$ is the minimum radius of the HYP, reached when u=0, while $\sqrt{r^2 + c^2} = R_{max}$ is the maximum radius, reached twice, when $u = \pm r \operatorname{arcsinh}(r/c)$. Visualizations of the HYP surface are shown in fig. 1.

It is important to notice that there are always values of u for which every pseudosphere, HYP, BEL and ELL, reaches its boundaries, corresponding to $R_{max}$. For BEL and ELL this happens once, while for HYP this happens twice. There $\frac{dR}{dz}$ reaches $\pm\infty$, the structure goes flat, hence the surface abruptly ends there. If we keep going, we merge into a plane, on both sides. Such boundaries are necessarily present, due to the Hilbert theorem on the embedding in Euclidean space [9]. In [4] these boundaries are named "Hilbert horizons", even though, in general, they are not general relativistic horizons. As mentioned, HYP, with its even function for $R(u)$, has two such 'horizons', along with all other values of the function (4), with the exception of $R = c = R_{min}$. In the Weyl symmetry approach to the carbon membranes - quantum gravity correspondence [1,5,8], those "Hilbert horizons", $\sqrt{r^2 + c^2}$, merge with the BH horizon only in the limit of $c/r \to 0$ [4]. In that limit HYP turns into two BELs joined by the tails (see (B) in fig. 1) and the BH it corresponds to has zero mass, $M = 0$ [4]. Therefore, all the difficulties of realizing BEL in a laboratory would actually be doubled. As recalled earlier, in the time-slice approach to the correspondence, it is $R = c$ where the BH horizon is located, henceforth no limit to an infinitely large carbon structure is necessary.

a          b

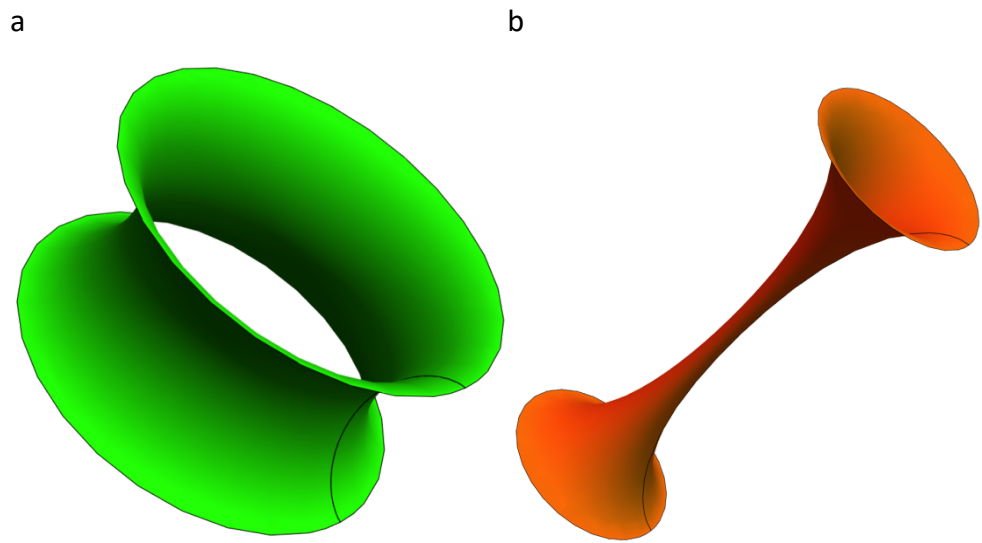

**Fig. 1.** HYP surfaces with parameters (a) r = 1 and c = 1 and (b) r =1 and c = 1/10

In our work we created a HYP with parameters r = 1 and c = 1 and it was scaled up by a factor of 20. This resulted in a HYP with an axis length of 28.5 Å and maximum and minimum diameters of 56.5 Å and 40 Å.

3 Computational details

Molecular Dynamics (MD) simulations were carried out with the LAMMPS open source code [15]. C-C interaction was described by the second generation reactive empirical bond order (REBO2) potential [16]. The REBO2 potential was constructed to reproduce the forming and breaking of bonds and is therefore well suited to describe situations involving bond rearrangements, such as reactions between and diffusion of point defects in graphene. HYPs with dimensions of some tens of Å were constructed and then embedded in graphene sheets of ~100 x 100 Å$^2$. The total number of atoms was always less than 10000. Such small systems do not scale efficiently over large numbers of cpu cores, hence calculations using mpi parallelization were run on no more than 8 cpu cores. A simulation of 100 million MD steps took approximately one week to complete.

The nano-extrusion and annealing simulations make extensive use of the 'walls' feature in LAMMPS, i.e. surfaces of various shapes (blocks, cones, cone slices, cylinders, ellipsoids, prisms, spheres, and also intersections or unions of such shapes) can be defined to which atoms experience a user-defined interaction as a function of distance.

We approximated the mathematical HYP surface by merging 88 slices of cones with gradually varying generatrix angles into a single body. While this is a convenient way to approximate a HYP with its rotational symmetry, it is of course possible to approximate almost any surface, including ones that do not have any symmetry and for which no analytical description Is available, by merging numerous smaller and bigger blocks into a single body. This means that the nano-extrusion process can be used to give almost any body a covering of carbon atoms in approximately the right areal density. The possible interactions that walls have with atoms include Lennard-Jones 12-6 (LJ) and repulsive harmonic potentials. The LJ wall used in our extrusion simulation is very strong, with a potential energy minimum of -20 eV. This (unrealistically strong) LJ wall merely serves to temporarily force carbon atoms to remain almost exactly on the mathematical HYP surface during extrusion and annealing. As a result, during annealing the point defects naturally position themselves such that the shape dictated by the LJ wall becomes a low stress, low energy, stable state for the graphene. When after 'soft-annealing' the LJ wall is switched off, the graphene retains a shape close to the LJ wall that was active during extrusion and annealing. The harmonic wall prevents carbon atoms from escaping well beyond the interaction range of the LJ wall, i.e. it guides carbon atoms towards the LJ wall.

4 Results

4.1 Nano-extrusion of thin carbon shapes

Fig. 2 shows the system used and Supplemental Material A shows an animation of the extrusion of the HYP. Supplemental Material B contains the LAMMPS input file of the extrusion simulation and configurations from the final part of the simulation. The LAMMPS input file can be used as a starting point for adapting and extruding different shapes. Supplemental Material B also contains input and output files for other types of simulations conducted in our work.

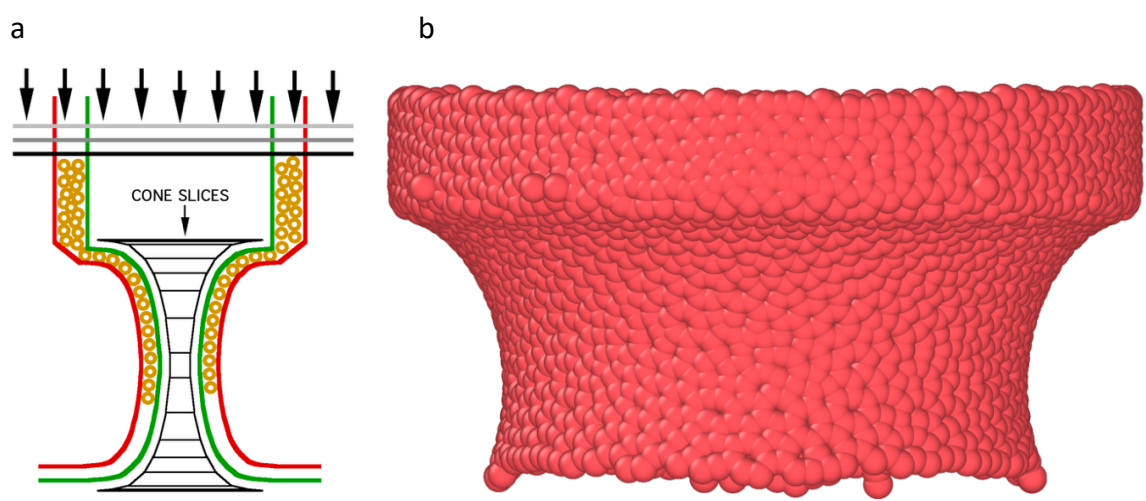

**Fig. 2.** a: schematic overview of the simulated HYP system. Green lines (smaller radius) represent a strong LJ wall, red lines (larger radius) represent a repulsive harmonic wall, the black line represents a harmonic wall that moves downward and thus acts as an extrusion press. b: snapshot from the HYP nano-extrusion simulation.

In addition to the use of unphysical walls, the extrusion simulation is less than fully realistic in some other ways. The extrusion press exerts very high pressure on the carbon that is initially placed in a hollow, thin cylinder volume above the HYP surface. The high pressure crushes the carbon structure and carbon flows between the LJ and harmonic walls like a strongly non-Newtonian fluid, offering significant resistance to flowing through its internal friction. This friction within the carbon limits the size of HYPs that can be produced, as the pressure exerted by the extrusion press is more and more undone the further carbon is ahead of the extrusion press. The pressure gradient also causes a slight gradient in carbon density and cohesion energy. The internal friction of the 'flowing' carbon imposes one limitation on the geometry that carbon can be extruded into. Narrow bottlenecks, followed by a much wider geometry should be avoided because the internal friction in the wider part would require extreme pressures to be exerted in the narrow part. This can lead to numerical instabilities where an atom is 'pushed through' the LJ wall, at which point the simulation halts with an error. In the HYP extruded in this work, the minimum/maximum radius ratio is 1 : 1.41, which was sufficiently low. For the extrusion of a HYP as in Fig. 1B, carbon would have to be injected from both ends, as the HYP is too narrow in the middle. Even when carbon extrusion is successful, the extruded carbon is in an unrealistic, unstable state. It is mostly made up of a mixture of pentagons, hexagons and heptagons and hardly anywhere do adjacent hexagons form a distinguishable piece of graphene. The average potential energy of the carbon atoms is several eV less than the cohesion energy of carbon atoms in diamond or graphite. The nano-extrusion method is suitable for occupying a HYP surface with a thin carbon structure, with an atomic surface density approximately that of graphene, but the highly defective resulting structure is unrealistically unstable if the LJ wall is switched off. The extruded structure needs to be transformed into realistic, stable graphene, such as would form in CVD experiments.

4.2 Relaxation and annealing of the extruded HYP

In order to turn the unstable extruded carbon into a stable graphene HYP, we relaxed and annealed the extruded structure, which has 2241 atoms. The harmonic outer wall and extrusion press wall were switched off, while the HYP-shaped LJ wall was kept on. As soon as the extrusion press is switched off, the carbon structure expands somewhat, slightly more so in the upper half than in the lower half, due to the density gradient that appeared during extrusion. Relaxation of the extruded carbon structure did stabilize it by 1.2 eV/atom, but this stabilization was mostly due to relief of the strong compressive stress after switching off the extrusion press wall, not due to the elimination of most of the point defects and the growth of mostly defect-free graphene. After the initial relaxation the structure was first annealed at 5 K with a very small timestep of 1 as. Subsequently it was annealed at higher and higher temperatures (up to 3500 K, to encourage fast kinetics while maintaining material integrity and MD simulation stability), with larger and larger timesteps (up to 1 fs for 3000 K), with a lower and lower energy value for the LJ wall (down to 0.2 eV), for longer and longer times (up to 0.1 µs). This annealing allowed most defects to rearrange and eliminate with 'opposite' point defects (e.g. two pentagons with two heptagons) as is seen in experiments, see e.g. [17], or become mobile and eliminate at the HYP Hilbert horizons (see

Section 2 and, e.g., [4]) that, in this configuration, coincide with the free edges of the structure. Also, in areas with an excess density of atoms and high compressive stress, a few atoms were ejected from the graphene and turned into adatoms on the graphene surface, thereby relieving the compressive stress. Both through the ejection of atoms from areas with strong compressive stress and expansion of the carbon structure, the density gradient present immediately after the extrusion process disappeared. As most of the defects disappeared, increasingly large areas of graphene developed. The total intrinsic curvature in the rearranging structure fluctuated, but developed towards the expected value of $K_{tot} = -4\pi$. Fig. 3 shows structures that were relaxed, with the HYP-shaped LJ wall switched off, before and after annealing for 30 ns at 3500 K. The initial structure had already undergone short annealing at temperatures up to 3000 K.

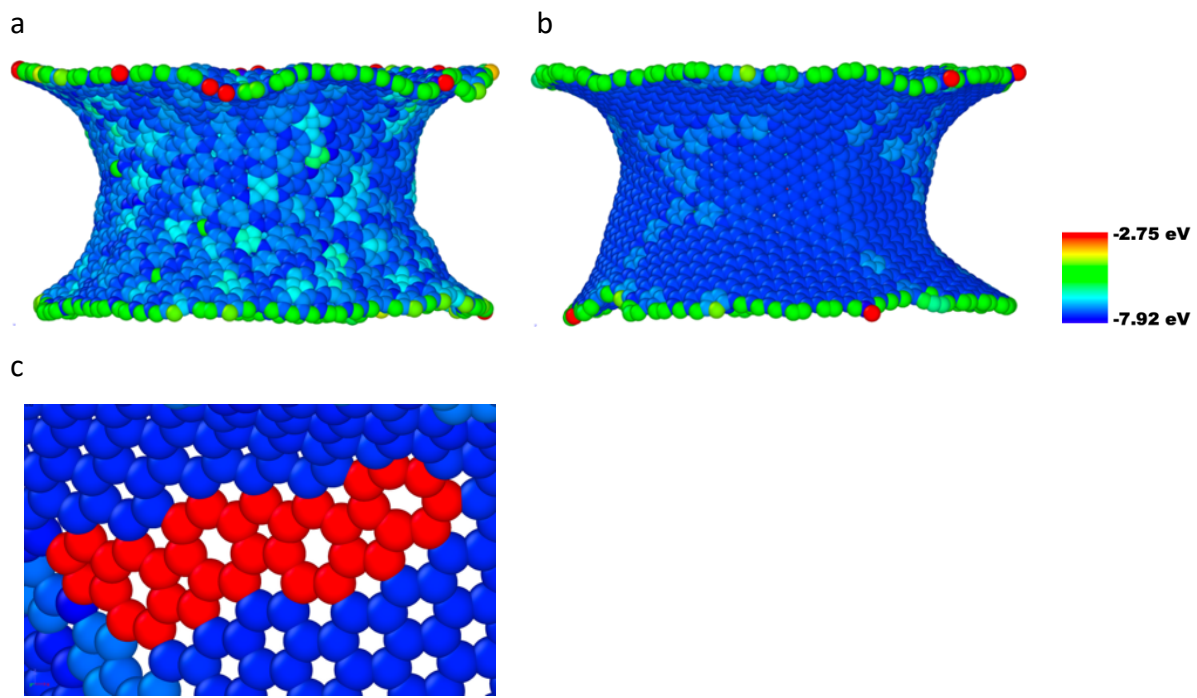

**Fig. 3.** Carbon HYP surface (a) before and (b) after 30 ns of annealing at 3500 K. Atoms in (a) and (b) are coloured by their potential energy, highlighting HYP Hilbert horizons, that here coincide with the free edges, and some types of point defects. (c) zoomed in portion of (b) with atoms inside a heptagon-pentagon grain boundary highlighted in red.

The structure in fig. 3B consists of polycrystalline graphene. The grain boundaries consist mostly of chains of alternating heptagons and pentagons. Grain boundaries may begin and end with a heptagon, thereby constituting one excess heptagon. In the system in fig. 3B, there were more than 50 heptagon-pentagon pairs (most of those in grain boundary chains) and three excess heptagons. Additionally, we observed a larger, more complex quad vacancy defect, consisting of a short chain of three asymmetric octagons with a pentagon at each end of the chain, see fig. 4A. The overall defect density in the HYP in fig. 3b is one defect per ~150 Å$^2$. This is an order of magnitude higher than the minimum required number of defects in the BEL constructed in [6].

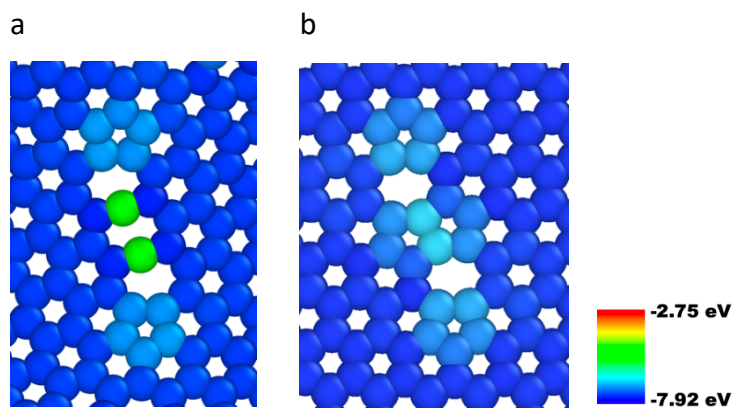

**Fig. 4.** Quad vacancy defect clusters (a) Cluster consisting of three octagons with a pentagon at each end (b) Two 5-8-5 defects that share two atoms in the middle. Atom are coloured according to their potential energy

The defect cluster can easily switch between the states in figs. 4A and 4B and this happened several times during the annealing at 3500 K. When the defect cluster in fig. 4A is transplanted into flat a graphene sheet and then annealed, it quickly transforms into the state in fig. 4B, which has a 0.74 eV lower energy. However, as demonstrated for carbon nanotubes [18], defect formation energies depend on the curvature of the graphene in which the clusters are embedded. Embedding the two clusters in fig. 4 in a carbon nanotube with a 4.1 Å radius reduces the energy difference to 0.49 eV. The curvature of the HYP and temperature effects at 3500 K cause the quad vacancy cluster to assume the complex form of fig. 4A part of the time. This demonstrates that while simple point defects [19-21] cause initially flat graphene to curve, once curved the defects may assume different, more complex states.

4.3 Attaching graphene sheets

While heptagon-pentagon pairs do not have a net effect on the graphene curvature, excess heptagons, the large defect in fig. 4A and the free edges have a strong effect. If after annealing with the LJ wall switched on the structure is then relaxed without the LJ wall, it deforms to a structure that deviates considerably from the mathematical HYP surface, see e. g. fig. 3B.

The distorting effect of the free edges on the carbon HYP can be removed by embedding the HYP in two periodic flat graphene sheets, see fig. 5. In this configuration, then, there are no free edges anymore, whereas the Hilbert horizons are the *loci* where the HYP ends, not the actual material edges of the structure.

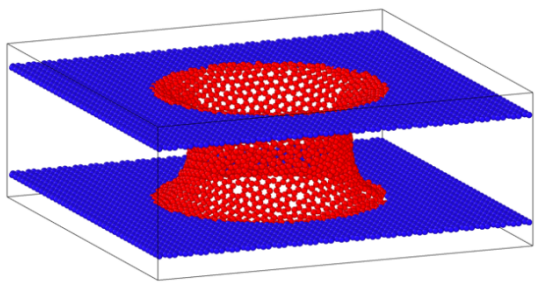

**Fig. 5.** Initial configuration of a HYP embedded in ~100 x 100 Å$^2$ flat graphene sheets

Since the HYP consists of polycrystalline graphene, it cannot fit coherently to the flat graphene sheets everywhere. Instead, we cut away the HYP atoms beyond a certain radius and created a circular hole in the graphene sheets with a 0.2 Å bigger radius and then fit the graphene sheets on the HYP. Due to the incoherent fit, this resulted in some sheet and HYP atoms initially being only a little more than 0.2 Å away from each other in some places, while there were small open gaps in other places. We then relaxed and annealed the combined system in stages, with the LJ wall switched on, as with the HYP.

Relaxing snapshots of the system at various stages of annealing without the LJ wall led to structures that deviated more strongly from the mathematical HYP surface than the initial structure. This turned out to be due to the sealing up of the small open gaps in the initial structure. Atoms across the open gaps want to form bonds and they will move towards each other to achieve this. This reduces the number of atoms with dangling bonds, but it creates tensile stress in the graphene. The tensile stress causes the graphene not to follow the mathematical HYP surface but to assume a shape where less graphene area is required, see fig. 6.

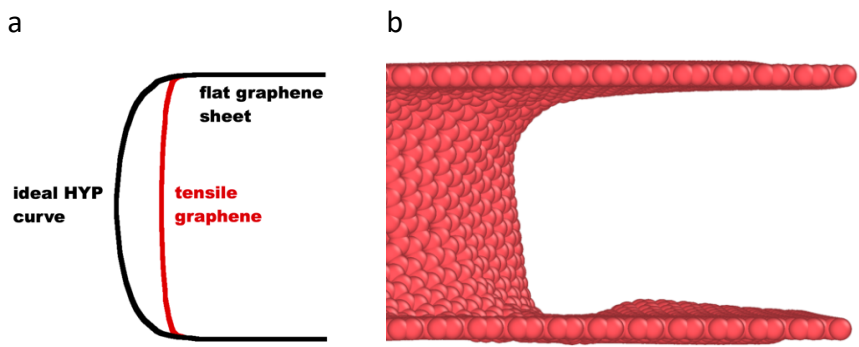

**Fig. 6.** Graphene under tensile stress pulling away from the mathematical HYP surface (a) Schematic overview (b) Relaxed configuration

A simple way to avoid graphene pulling loose from the mathematical HYP surface is to make the radius of the hole in the flat graphene sheets slightly smaller than the radius of the HYP. That way there will be fewer and smaller open gaps. There will also be more places in the initial structure where HYP and sheet atoms are very close together, but during annealing at high temperature these excess atoms get ejected from the graphene as adatoms, giving a relatively smooth connection between the HYP and the graphene sheets. Annealing also helps to reduce residual stresses to low levels, thus ensuring there is little interaction between periodic HYP images across the ~100 Å in-plane box lengths. Fig. 7 shows the top view of the initial system and the system at various stages of annealing.

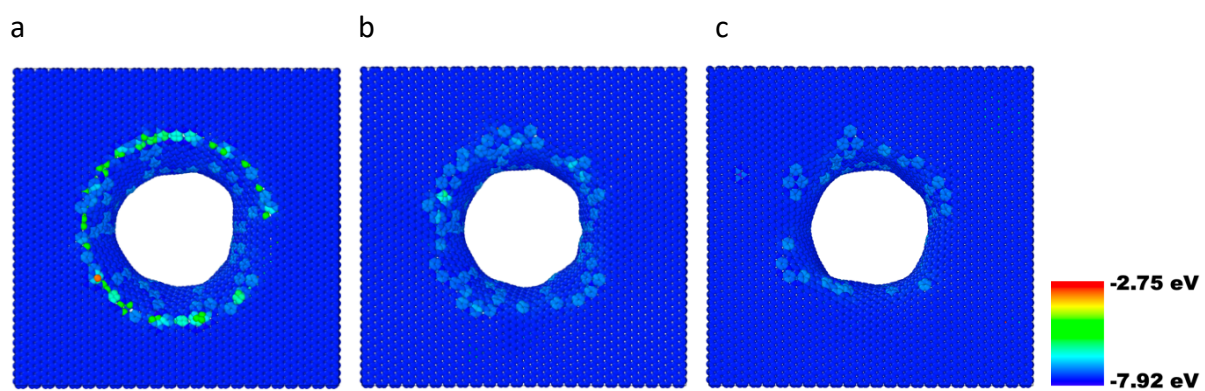

**Fig. 7.** Top view of HYP with flat periodic graphene sheets attached (a) Relaxed initial system (b) System after annealing for 5 ns at 3000 K (c) System after annealing for 0.1 μs at 3000 K. Atom are coloured according to their potential energy.

With a smooth connection between the HYP and graphene sheets, and Hilbert horizons no longer ending in free graphene edges, carbon atoms assume positions relatively close to the mathematical HYP surface. Fig. 8A shows an annealed and then relaxed structure, with no LJ wall to artificially keep the HYP in shape. The root means square (rms) distance between carbon atoms inside the HYP and the mathematical HYP surface is 0.60 Å. The deviation between graphene atoms and the mathematical HYP surface is largest near the sheets-HYP connecting rings.

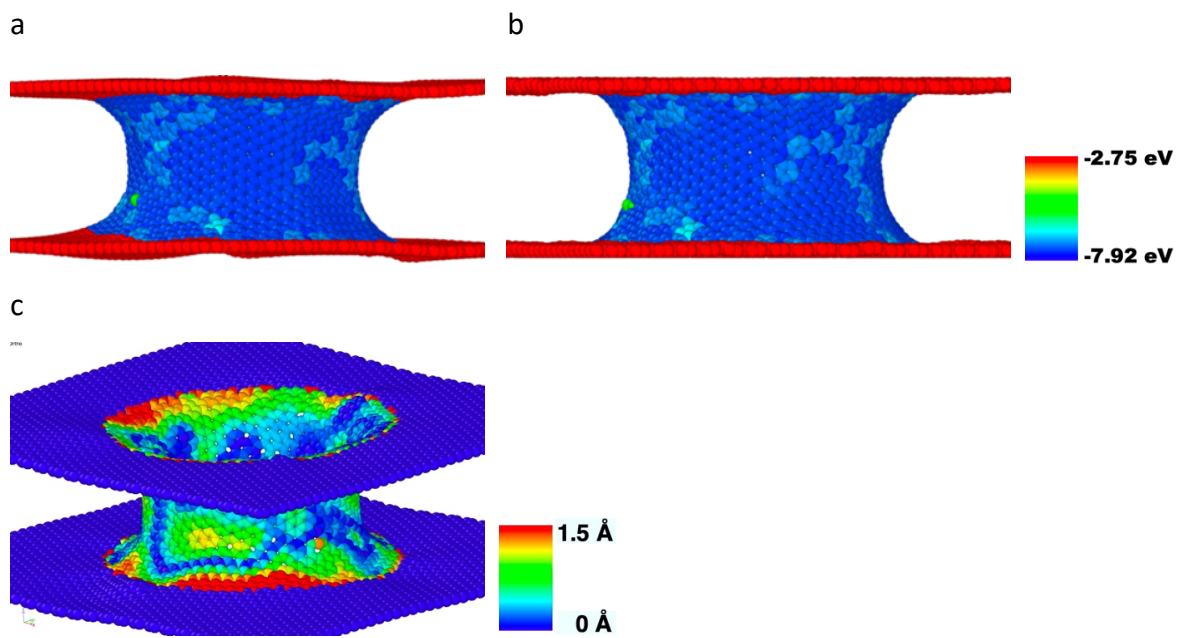

**Fig. 8.** Annealed and relaxed HYP with attached flat graphene sheets (a) Without any active walls in the system (b) With weak, flat LJ walls active parallel to the flat graphene sheets. In (a) and (b), the atoms outside the HYP are coloured red (top and bottom atomic planes), atoms inside the HYP (middle) are coloured according to their potential energy. (c) The annealed and relaxed HYP of (a), coloured by deviation between the graphene HYP atoms and the mathematical HYP surface.

The initially flat graphene sheet in fig. 8A has been visibly deformed during relaxation, i. e. the HYP portion of the system shows a tendency to deform and the flat graphene sheets only partly restrain this deformation. Adding extra flat graphene sheets at both the top and bottom of the system would restrain the deformation more strongly because the areal moment of inertia of the graphene sheets goes up with the cube of the thickness of the stack of sheets. We investigated the limiting case of adding many sheets at the top and bottom by activating LW walls parallel to the flat graphene sheets (with holes in them the size of the HYP, so that they only cover the flat graphene sheets). This forces the graphene sheets to remain almost perfectly flat, as if there are many graphene sheets attached. The effect of such a strong restraint is to keep the HYP atoms much closer to the mathematical HYP surface, see fig. 8B. The rms of the distance between carbon atoms and the HYP surface is reduced to 0.20 Å. While limiting the relaxation leads to a structure that is closer to the mathematical HYP surface, it produces a less stable system. The formation energy of the fully relaxed system (i. e. the potential energy of the HYP system with graphene sheets relative to the potential energy of a flat, defect-free graphene sheet with an equal number of atoms) is 300 eV, while for the HYP system with flat graphene sheets it is 439 eV. In a system that was created with slightly different initial conditions but underwent the same annealing and relaxation, the rms deviations for the systems without any LJ walls and with a LJ wall to keep the graphene sheets flat were 0.54 and 0.22 Å and the formation energies were 235 and 370 eV.

When a HYP is used as the spatial part of a BH continuum metric, as seen by graphene conductivity electrons, the electron wavelength of interest is larger than the C-C distance of graphene, $\lambda > 1.4$ Å. In that regime it is the case that both the electrons are well described by a Dirac field and that the lattice is well approximated by a continuous manifold [7]. Deviations of 0.20 Å are then acceptable. In fact, the focus of gravity analog experiments is on regimes where the wavelength is actually comparable to the radius of curvature [4], r, that for us means $\lambda > 40$ Å.

4.4 Structural stability

Since the goal of the simulations was to test if graphene HYPs would be stable enough to conduct experiments on them, we rigorously tested their mechanical stability. We imposed large uniform deformations and then let the system evolve in short (0.1 ns) constant energy simulations. Relaxed systems were elongated 5, 10 and 20% along the HYP axis and sheared 5, 10 and 20° perpendicular to the HYP axis. Strongly deforming the systems and then releasing them leads to cycles of compression-expansion or shearing back and forth with a gradually fading amplitude. Also, there were smaller wavelength ripples temporarily present in which the graphene is strongly bent and has a small radius of curvature. This is a very demanding test of the mechanical stability of the HYPs. Supplemental Material C [17] shows animations of the strongly deformed HYPs after being released. As time progresses, the initial strain energy partly dissipates into heat. In the systems that were elongated/sheared by 20 %/° the temperature rose 364 and 265 K, respectively. After the dynamics simulations, systems were relaxed again.

All elongated systems returned to the exact same state after 0.1 ns of MD and then relaxation as before the MD, as did the system that was sheared 10°. The systems that were sheared 5 and 20° relaxed towards a slightly different state after the MD, in which a few clustered point defects had rearranged. The system energy of the new state was 0.05 eV higher than the original state, i.e. the structural transformation induced by the shearing and then release is small and energetically insignificant. Overall, the HYP systems with flat graphene sheets attached show considerable stability and flexibility under large externally imposed deformations. They also show full recovery after the imposed deformation is removed.

In addition to imposing mechanical deformations, the structural stability of the HYPs was also tested by annealing for 0.1 μs at 500, 1000 and 1500 K without any LJ walls or other artificial constraints. This annealing did not lead to any structural transformation.

The mechanical testing and annealing shows that the graphene HYPs are quite stable. Once created in experiments, they will likely persist for macroscopic times and be robust enough to conduct experiments on them.

## 5 Summary and future prospects

We presented a new simulation method for creating stable curved graphene surfaces. Classical MD simulations were used to perform a nano-extrusion process in which carbon atoms were forced down a thin three-dimensional volume. The volume can take almost any shape and does not require any symmetry or an analytical description. Extrusion of carbon requires a very high pressure that turns the carbon into an unrealistic state several eV/atom less stable than graphite or diamond. The unstable carbon does have an areal density close to that of graphene. Annealing with a constraint to keep the carbon in the desired shape turns the carbon into realistic polycrystalline graphene. Our method offers a relatively uncomplicated way of producing simulated graphene structures in a wide variety of possible shapes. The method can be used to e.g. test if graphene structures that are to be produced experimentally are stable.

Motivated by the possibility to reproduce analog quantum gravitational phenomena, such as Hawking effect, on carbon-made membranes, we have applied our new method to the creation of carbon HYPs. Nano-extrusion created a carbon HYP precursor that was unrealistically unstable, but did have approximately the correct areal carbon density. Subsequent annealing made the carbon precursor crystallize into polycrystalline graphene with a strongly reduced number of lattice defects. The number of excess heptagons fluctuates during annealing and they cause a total Gaussian curvature that develops towards $K_{tot} = -4\pi$. The point defects keep the graphene broadly in the HYP shape when the constraint is removed, though the free pseudosphere edges cause considerable deviation. The deforming influence of free edges can be removed by embedding the pseudosphere in flat graphene sheets with periodic boundaries. The initial fit between flat graphene and the polycrystalline curved graphene of the hyperbolic pseudosphere is always mostly incoherent and introduces strong stresses around the connecting circles. These stresses can be removed by annealing again with constraints in place to keep both the HYP and flat graphene sheets in their shapes. This leads to a smooth connection with relatively few defects between the HYP and the flat graphene sheets. When the constraints are removed after annealing, the carbon atoms in the HYP part of the structure (which in our case had dimensions of some dozens of Å) deviate from the ideal mathematical HYP with a rms of 0.6 Å. When the constraint that keeps the periodic graphene sheets flat is kept in place, as if instead of single graphene sheets there are stacks of many sheets with high stiffness, the rms deviation of carbon atoms in the hyperbolic pseudosphere reduces to 0.2 Å.

The focus of gravity analog experiments is on regimes where graphene quasiparticles' wavelength has the radius of curvature as lower bound. In the case we have studied here, then, it must be λ > r = 40 Å, that means typical energies of the quasi particles E < 0.1 eV. Although the precise calculations will only become available when the full analog gravity model will be developed, based on earlier work [4] we can estimate that deviations from the ideal structure of up to 10 per cent of r, that here means 4 Å, are the limiting value to still have some Hawking-like signals in the local electronic density of states (LDOS).

What we refer to here is the analysis of [4] of the non-thermal corrections to the LDOS due to deviations from the ideal BEL. There a composite parameter b, related to the radius of curvature r, is introduced to measure how far from (Hawking) thermal behaviour of the LDOS one gets, when measurements are taken far from the horizon and close to the non-ideal end of BEL [4]. When such b is infinite (1/b is zero) the structure is ideal. When b is zero (1/b is infinite) the Hawking thermality is fully gone. From the analysis there, a value of b = 10 (that is, 1/b is ten per cent) gives a signal to noise ratio at the limit. From there originate our statements above.

Again, given that the theoretical models are different (Weyl symmetry for [4], time-slicing here) and so are the carbon pseudospheres (BEL for [4], HYP here), this is only a rough estimate. Nonetheless, the results of this paper of a deviation of 0.60 Å, that is only a 15 per thousand portion, is quite remarkable and reassuring for future analog gravity experiments.

After annealing, the structures with periodic graphene sheets were subjected to rigorous tests of their mechanical stability. The structures proved stable against externally imposed 20% elongation or 20°shearing and then release, as well as against annealing at 1500 K. This shows that HYP produced through experimental techniques will likely be robust enough to then conduct experiments on them.

While in this work we only made carbon structures through nano-extrusion and then annealing, the method may be applicable to other (combinations of) elements that form monolayer structures, like boron nitride. In future work we plan to investigate if other simulated curved monolayer materials can be produced in the same way.

Acknowledgements

T. P. C. K thanks Jan Steven Van Dokkum for running some of the long calculations. A.I. and V.T. thank Raffaele Agostino, Andrea Legramandi, Pavel Krtouš and Petr Lukeš for valuable discussions and gladly acknowledge support from Charles University Research Center (UNCE 24/SCI/016). The work of D. L. and T. P. C. K. was supported by e-INFRA project CZ(ID:90254).

Data availability

Supplemental Material B contains input and output files for different types of simulations conducted in this work, i.e. extrusion, annealing of a HYP, annealing of a HYP with periodic flat graphene sheets attached and mechanical testing.

CRediT author statement

**T. P. C. Klaver**: Data curation, Formal analysis, Investigation, Methodology, Software, Validation, Visualization, Writing - original draft, Writing - review & editing. R**. Gabbrielli:** Formal analysis, Investigation, Software, Visualization. **V. Tynianska**: Formal analysis, Investigation. **A. Iorio**: Conceptualization, Formal analysis, Funding acquisition, Investigation, Project administration, Supervision, Writing - review & editing. **D. Legut**: Investigation, Funding Acquisition, Supervision, Writing - review & editing.

Statements and Declarations

The authors declare no special interests.

References

[1] A. Iorio, "Weyl-gauge symmetry of graphene", Ann. Phys. 326 (2011) 1334-1353
https://doi.org/10.1016/j.aop.2011.01.001

[2]   A. Iorio, "Using Weyl symmetry to make graphene a real lab for fundamental physics" Eur. Phys. J. Plus, 127 (2012) 156  https://doi.org/10.1140/epjp/i2012-12156-1

[3]   A. Iorio, G. Lambiase, Physics Letters B, "The Hawking–Unruh phenomenon on graphene", 716 (2012) 334-337  https://doi.org/10.1016/j.physletb.2012.08.023

[4]   A. Iorio and G. Lambiase, "Quantum field theory in curved graphene spacetimes, Lobachevsky geometry, Weyl symmetry, Hawking effect, and all that", Phys. Rev. D 90 (2014) 025006  https://doi.org/10.1103/PhysRevD.90.025006

[5]   A. Iorio, "Curved spacetimes and curved graphene: A status report of the Weyl symmetry approach", Int. J. Mod. Phys. D 24 (2015) 1530013 http://dx.doi.org/10.1142/S021827181530013X

[6]   Simone Taioli, Ruggero Gabbrielli, Stefano Simonucci, Nicola Maria Pugno, Alfredo Iorio, "Lobachevsky crystallography made real through carbon pseudospheres", J. Phys.: Condens. Matter 28 (2016) 13LT01  https://doi.org/10.1088/0953-8984/28/13/13LT01

[7]   A. H. Castro Neto, F. Guinea, N. M. R. Peres, K. S. Novoselov, A. K. Geim, "The electronic properties of graphene", Rev. Mod. Phys. 81 (2009) 109 https://doi.org/10.1103/RevModPhys.81.109

[8]   G. Acquaviva, A. Iorio, P. Pais, L. Smaldone, "Hunting Quantum Gravity with Analogs: the case of graphene", Universe, 8 (2022) 455 https://doi.org/10.3390/universe8090455

[9]   L. P. Eisenhart, "A Treatise on the Differential Geometry of Curves and Surfaces", Princeton University Press, Princeton, NJ, 1909

[10]  A. Ovchinnikov, "Gallery of pseudospherical surfaces, in Nonlinearity and Geometry", edited by D. Wojcik and J. Cieslinnski, Polish Scientific Publishers PWN, Warsaw, 1998, 41–60

[11]  V. Tynianska, A. Iorio, in progress

[12]  V. Tynianska, R. Gabbrielli, T. P. C. Klaver, D. Legut, A. Iorio, in progress

[13]  M. Bañados, C. Teitelboim, J. Zanelli, "The Black Hole in Three-Dimensional Spacetime", Phys. Rev. Lett. 69 (1992) 1849–1851 https://doi.org/10.1103/PhysRevLett.69.1849

[14]   V. P. Kurupath, S. K. Kannam, R. Hartkamp, S. P. Sathian, "Highly efficient water desalination through hourglass shaped carbon nanopores", Desalination 505 (2021) 114978  https://doi.org/10.1016/j.desal.2021.114978

[15]   S. Plimpton, "Fast Parallel Algorithms for Short-Range Molecular Dynamics", J. Comput. Phys. 117 (1995) 1  https://doi.org/10.1006/jcph.1995.1039

[16]   D. W. Brenner, O. A. Shenderova, J. A. Harrison, S. J. Stuart, B. Ni, S. B. Sinnott, "A second-generation reactive empirical bond order (REBO) potential energy expression for hydrocarbons", J. Phys.: Condens. Matter 14 (2002) 783–802 https://doi.org/10.1088/0953-8984/14/4/312

[17]   O. Dyck, S. Yeom, S. Dillender, A. R. Lupini, M. Yoon, S. Jesse, "The role of temperature on defect diffusion and nanoscale patterning in graphene", Carbon 201 (2023) 212–221  https://doi.org/10.1016/j.carbon.2022.09.006

[18]   J. M. Carlsson, "Curvature and chirality dependence of the properties of point defects in nanotubes", Phys. Stat. Sol. (b) 243 (2006) 3452–3457 https://doi.org/10.1002/pssb.200669223

[19]   L. Liu, M. Qing, Y. Wang, S. Chen, "Defects in Graphene: Generation, Healing, and Their Effects on the Properties of Graphene: A Review", J. Mater. Sci. Technol. 31 (2015) 599-606  https://doi.org/10.1016/j.jmst.2014.11.019

[20]   P. T. Araujo, M. Terrones, M. S. Dresselhaus, "Defects and impurities in graphene-like materials", Mater. Today 15 (2012) 98-109 https://doi.org/10.1016/S1369-7021(12)70045-7

[21]   W. Tian, W. Li, W. Yu, X. Liu, "A Review on Lattice Defects in Graphene: Types, Generation, Effects and Regulation", Micromachines 8 (2017) 163 https://doi.org/10.3390/mi8050163